\documentclass[12pt,oneside,onecolumn]{elsart}

\usepackage[nospace]{cite}
\usepackage[latin1]{inputenc}
\usepackage{umlaute}
\usepackage[english]{babel}
\usepackage{amssymb,amstext,amsfonts}
\usepackage[intlimits]{amsmath}
\usepackage{graphicx,epsfig}
\usepackage{latexsym,layout}
\usepackage{lineno}

\inputencoding{latin1}

\begin{document}

\begin{frontmatter}
 
\title{Simulated performance of a single pixel PIN spectrometer SCXM equipped with a 
       concentrator optics in Solar coronal X-ray observations}

\author[Astronomy]{L. Alha\thanksref{Principal}},
\author[Astronomy]{J. Huovelin},
\author[FINCA]{J. Nevalainen$^{a, }$}

\address[Astronomy]{Department of Physics, Division of Geophysics and Astronomy, P.O. 
Box 48, FI-00014 University of Helsinki, Finland}
\address[FINCA]{Finnish Centre for Astronomy with ESO, University of Turku, V\"ais\"al\"antie 20, 
FI-21500 Piikki\"o, Finland \& Department of Physics, Division of Geophysics and Astronomy, P.O. Box 48, 
FI-00014 University of Helsinki, Finland}

\thanks[Principal]{Corresponding author. E-Mail:  alha@mappi.helsinki.fi, Tel. +358-9-19151672}

\begin{keyword}
Off-limb coronal X-ray observations, solar axions
\end{keyword}

\end{frontmatter}


\vspace{0.0cm}

\begin{frontmatter}

\begin{abstract}

In this paper we present simulated solar coronal X-ray observations to verify the sensitivity of
a new hypothetical instrument design. These simulations are folded through this X-ray spectrometer 
having a moderate size circular field of view ({\it FoV})) of 1.6$^{\circ}$. This SCXM (Solar Coronal 
X-ray Mapper) is designed to compose of a single pixel silicon PIN detector equipped with a 
single reflection double frustum X-ray optics. A moderate FoV would enable a morphological 
study of the expanded X-ray emission from the solar corona during a high activity of the Sun. 
The main scientific task of SCXM would be the mapping of the coronal X-ray emission, i.e. 
to resolve the radial distribution of the X-ray surface brightness around the Sun. These 
kind of off-limb observations would help to interpret the coronal plasma diagnostics as a 
function of the elongation angle.\\
Direct solar full disc observations could be also performed with SCXM. In this work we have applied 
real solar coronal X-ray data obtained by the SMART-1 XSM (X-ray Solar Monitor) \cite{Alha1} 
to simulate on-solar observations at different flux levels to derive full disc sensitivity and 
performance of SCXM.\\
A challenging attempt for SCXM would also be to distinguish the X-ray spectrum of the decaying 
axions around the Sun. These axions are assumed to be created as side products of fusion 
reactions in the core of the Sun. These axions are predicted to be gravitationally trapped to 
orbit the Sun forming a halo-like X-ray emitting object. No signature of an axion X-ray emission 
around the Sun has been observed to this day.\\
This simple X-ray spectrometer with an optical concentrator would be an inexpensive instrument 
with low mass and telemetry budgets compared with more accurate X-ray instruments of imaging 
capability. Hence SCXM would be an advanced choice as an auxiliary instrument for solar coronal 
X-ray observations.\\

\end{abstract}

\end{frontmatter}

\clearpage

\section{Introduction}
\label{Introduction}

The technical design of SCXM is introduced in \cite{Alha2} and the corresponding information is 
also compiled in Table 1. SCXM is designed mainly for off-limb X-ray observations. 
There are no on-going space missions dedicated solely to off-limb coronal X-ray observations. There 
have been a variety of solar EUV and X-ray missions carrying on-board grazing incidence or multilayer 
mirror telescopes having a small size FoV for direct solar full disc observations. There are several 
high energy solar missions operating also today, e.g. TRACE, RHESSI, Solar-B and STEREO, which all 
have an imaging instrument with a FoV less than 1$^{\circ}$. The YOKHOH SXT is the only solar X-ray 
instrument, which has made a few off-limb broad band spectroscopic observations \cite{YOKHOH}. 
SphinX instrument on-board the Russian CORONAS-PHOTON satellite has also made full limb solar coronal 
X-ray observations between 1.0 and 15.0 keV with single pixel silicon PIN diodes in 2009 
\cite{SPHINX1}.\\
\hspace*{0.5cm} Our SCXM would have about 10 times better energy resolution compared to the existing imaging solar 
X-ray instruments. The FoV of SCXM is optimized with respect to the moderate size of FoV versus its 
grasp {\it G}. Grasp denotes a product of an effective area {\it $A_{eff}$} and a FoV with a solid 
angle {\it $\Omega$}. The performance of grasp is related to instruments observing extended targets. 
Solar corona can be assumed to be a surface emitter of soft X-rays with several possible sources,
e.g. the solar coronal plasma or gravitationally trapped Kaluza-Klein axions decaying within the 
outer solar corona.\\
\hspace*{0.5cm} The detector concept of SXCM is a heritage from SMART-1 and Chandrayaan-1 XSMs \cite{Alha1}. The 
main difference is the applied double frustum single reflection optical concentrator in front of 
the detector PIN diode. The geometric active area of the silicon detector would be maximized so 
that the innermost guard ring of the silicon PIN detector diameter is 5.0 mm. The greater the 
active area, more on-axial X-ray photons can enter the detector without reflection. The 5.0 mm 
diameter is a practical trade-off value. A larger active detector area of a PIN diode would 
introduce a higher capacitance, which degrades the energy resolution. The aim is to keep energy 
resolution {\it $\Delta$E} of SCXM around 250 eV at 6 keV. The drawings of the SCXM optics are 
shown in Figs. 1 and 2. The dimensional design of SCXM is based on 
the optimization of the grasp {\it G}. The sensitivity of SCXM is optimized so that the integrated 
total grasp has its maximum between 1.5 keV and 10 keV.\\
\hspace*{0.5cm} There would be two different possible observing scenarios for SCXM. It should be located either 
on-board a rotating 1D-stabilized or three axis stabilized satellite platform. On board a spinning 
1D-stabilized platform SCXM would make observations in a rotational scanning mode, i.e. the FoV 
would sweep an annular area around the Sun in this off-limb mode. On board a 3D-stabilized platform 
SCXM would have a static and constant small off-limb elongation. These two possible observation 
strategies are shown in Fig 3. SCXM could be aligned also directly toward the Sun 
parallel with an imaging telescope in both modes. In this case SCXM could improve a combined 
observation with a better energy resolution compared to, e.g. a master imaging X-ray instrument.\\
\hspace*{0.5cm} The integration time would determine the observed annular target area during a off-limb 
rotational scanning mode. Hence the spinning rate of a satellite and the instrumental sensitivity 
would determine the azimuthal spatial resolution of the obtained data. During a long integration 
the annular target area would contain several revolutions and the azimuthal surface brightness 
information would be totally lost. In this case the observed data would include only the information 
of the radial distribution of the off solar X-ray spectral data.

\begin{figure}
   \begin{center}
    \epsfig{file=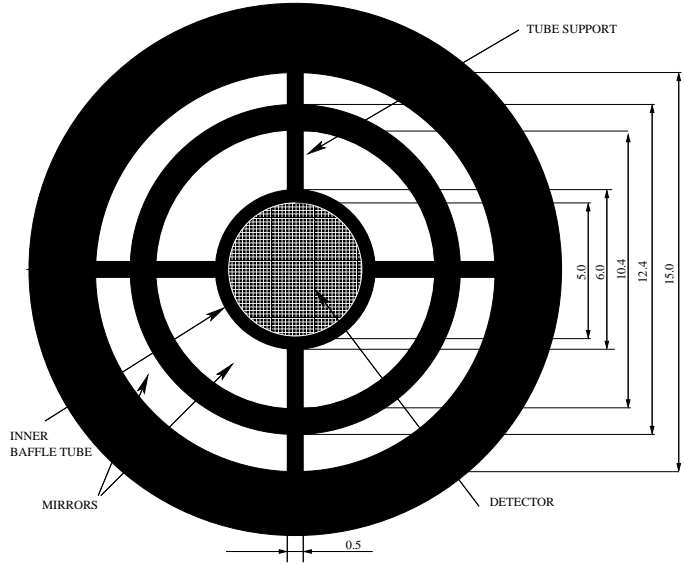,scale=0.60,keepaspectratio}
    \label{Fig:fig1}
    \end{center}
\caption{A dimensional drawing of the SCXM double frustum optics.}
\end{figure}

\begin{figure}
   \begin{center}
    \epsfig{file=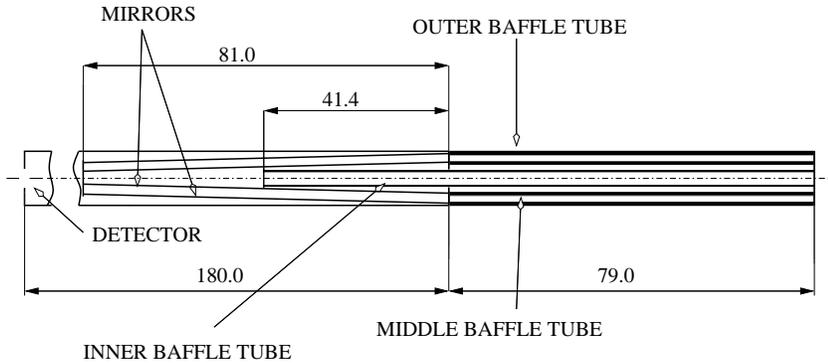,scale=0.60,keepaspectratio}
    \label{Fig:fig2}
    \end{center}
\caption{A front view illustration of the SCXM double frustum optics. The design of the SCXM optics is 
              introduced in \cite{Alha1}.}
\end{figure}

\begin{figure}
   \begin{center}
    \epsfig{file=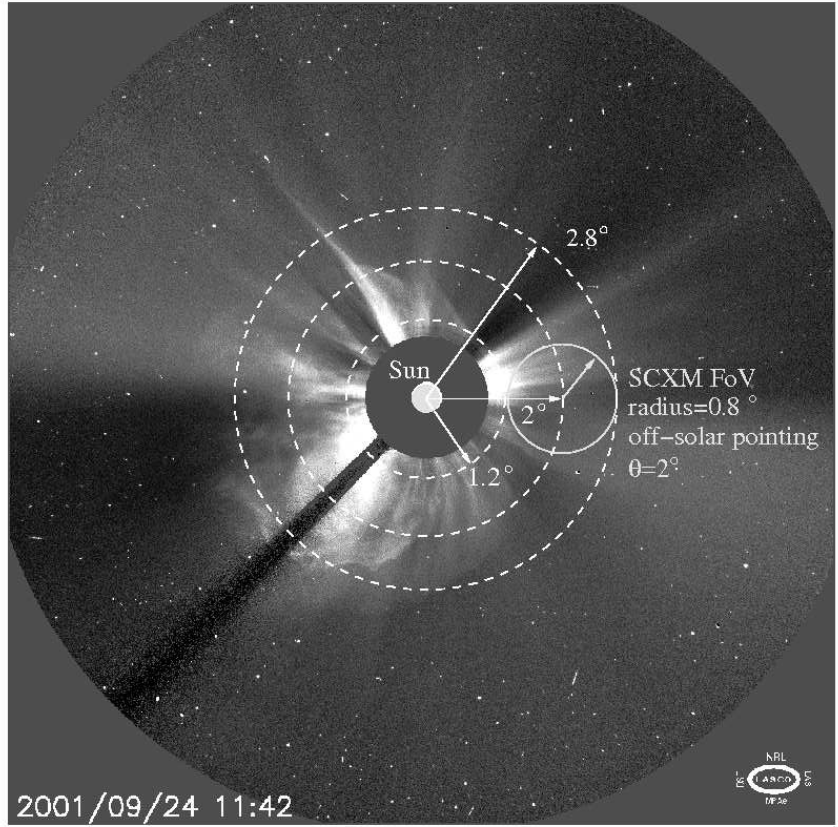,scale=0.55,keepaspectratio, angle=0}
    \label{Fig:fig3}
    \end{center}
\caption{This SOHO LASCO C3 white light coronal mass ejection (CME) image illustrates the FoV size
              of SCXM compared to the dynamical structures of the outer corona. The circle with a solid outline 
              illustrates a static off-solar pointing and the dashed white concentric circles outline the 
              rotational scanning mode. All CMEs do not coincide with flares with increasing X-ray emission. 
              An X2.6 flare peaked about 1 hour before this image was taken. An X2.6 flare corresponds to the 
              solar X-ray flux of $2.6 \times 10^{-4}$ Wm$^{-2}$ in the energy range of 1.55-12.4 keV (GOES standard) 
              \cite{GOES}. The solar X-ray flare classification is based on the logarithm of the flux in the 
              1.55-12.4 keV band, so that each flare is identified by a letter plus a numeric code indicating
              the subclass. Photo courtesy: NASA/SOHO.} 
\end{figure}

\begin{figure}
   \begin{center}
    \epsfig{file=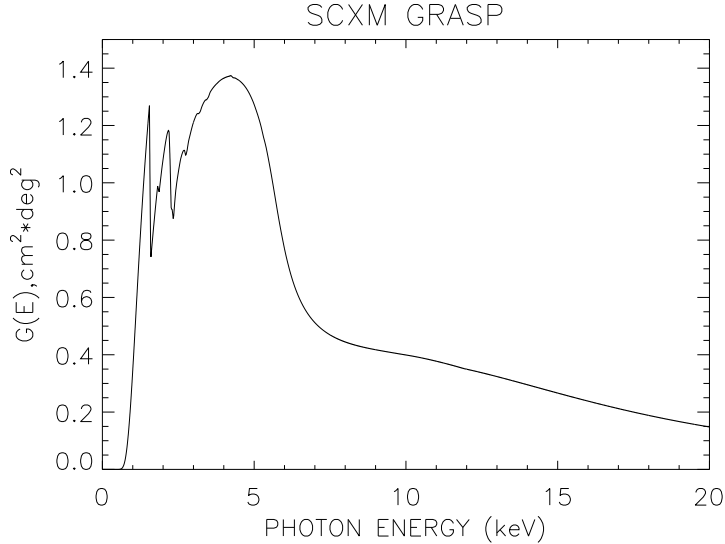,scale=0.60,keepaspectratio}
    \label{Fig:fig4}
    \end{center}
\caption{SCXM grasp as a function of photon energy}
\end{figure}

\section{Solar coronal X-ray emission targets of SCXM}
\label{Solar coronal X-ray emission targets of SCXM}

SCXM would be a non-imaging instrument. Hence SCXM would be concentrated to observe solar coronal physical 
phenomena, which do not require exact knowledge of a morphology, a size or absolute dimensions of the 
emitting target. The main objective would be observations of flare events covering a wide flux scale.
The solar flares originate from magnetic flux tubes penetrating out of the photosphere. 
The solar activity is a manifestation of the magnetic field originating from the solar convective zone. 
The magnetic fields are caused by the so-called solar dynamo. The magnetic dynamo is maintained and amplified
by differential rotation and turbulence in the outer convective zone of the Sun. A magnetic energy is 
converted into a heat during a flare event via a reconnection. Magnetic reconnection occurs, when a magnetic 
flux line interacts with a same magnetic flux line coming from the opposite direction. Practically a 
reconnection requires an magnetic loop to bend or to twist for touching itself above the photosphere. A 
magnetic energy is released in a wide band of electromagnetic radiation including also a soft X-ray emission. 
The magnitude of a flare event can be recorded by measuring the X-ray flux and its duration from an onset up 
to a decline phase. X-ray flare fluxes can vary several orders of magnitudes. The weakest observed solar 
coronal X-ray flux between 1.55 keV and 12.4 keV is about 10$^{-10}$ Wm$^{-2}$ \cite{SPHINX1} and the highest is 
about  10$^{-3}$ Wm$^{-2}$ at 1AU until now. Time scales of flares can vary also from several minutes up to 
several hours depending on the intensity of an event. An X-ray free-free emission of hot thin plasma can 
be modeled with an isothermal breaking radiation spectrum. The temperature and electron density changes 
during a flare event, which could be measured by SCXM. The measured flare temperatures can vary between 
1MK and 100MK. The emission measure describing the electron density of a thin hot solar coronal plasma typically 
varies between 10$^{46}$m$^{-3}$ and 10$^{50}$m$^{-3}$ \cite{Shibata}. The ionization stages of coronal elements change 
during a flare event due to the heating of the coronal plasma. The emission lines of highly ionized elements 
are clearly seen in the flare spectra above C-level, like Helium (FeXXVI) and Hydrogen (XXVII) like iron ions. 
The energy centroids of these lines help to interpret the X-ray data related to the plasma diagnostics of  
flare events.\\
\hspace*{0.5cm} Some CMEs also coincide with flares of enhanced X-ray emission. Plasmoids released by huge CMEs bring enormous 
amounts of hot X-ray emitting plasma in the outer solar corona. One of the primary tasks of SCXM would be to 
map this kind of X-ray emission from the expanded solar corona during a high activity in the off limb mode. 

\section{Full disc solar sensitivity simulations}
\label{Full disc solar sensitivity simulations}

The aim of this simulation was to determine the performance of SCXM with respect to the limiting 
count rate in on-solar, i.e. full disc observations. The background is negligible in full disc solar 
observations. Hence this simulation does not include effects of background emission.\\
\hspace*{0.5cm} The spectral data simulations require the knowledge of the total on-axis grasp ({\it G}) of the 
instrument (see Fig. 4). All the factors affecting the SCXM grasp are compiled in 
Table 1. This grasp is loaded into the Ancillary Response File ({\it ARF}) required 
in the spectral fitting with XSPEC S/W \cite{Arnaud}. The other standard input for the XSPEC is the 
Redistribution Matrix File ({\it RMF}). {\it RMF} includes also the information of the detector 
energy scale and the current energy resolution as a function of photon energy. The energy resolution 
of SCXM used in these simulation was 250eV @ 6keV. The nominal operational energy range of SCXM would 
be between 1-20 keV or optimally 1-10 keV.\\
\hspace*{0.5cm} We simulated measured count rates by folding four solar spectra through the SCXM response. Each 
spectrum corresponded to a different flux level. Those fluxes were B1 (1.0 $\times 10^{-7}$ Wm$^{-2}$), 
C2 (2.0 $\times 10^{-6}$ Wm$^{-2}$), C6 (6.0 $\times 10^{-6}$ Wm$^{-2}$) and X1 (1.0 $\times 10^{4}$ Wm$^{-2}$). The simulated spectral 
models were isothermal and they were composed of bremsstrahlung continuum and Gaussian lines 
(see Table 2.). The model parameters were derived from SMART-1 XSM data \cite{Alha1}. 
The plot of the simulated count rates are shown in Fig. 5. This simulation reveals 
that SCXM would be too fast or too sensitive for on-solar observations equipped with the filter 
configuration and optical dimensions listed earlier. There would be two limiting factors of the 
operation of SCXM with respect to the observed count rate. The first factor would be related to 
the pile-up counts. The lower horizontal line in the plot represents the pile-up limit of 3\% of 
the existing acquisition readout electronics. This 3\% pile-up corresponds to a flux B5.5 according 
to the GOES scale and it represents a quite low solar activity. The upper horizontal line represents 
the saturation limit of the existing XSM electronics. The filter transmission should be decreased or 
the optics should be reduced, if SCXM is used in on-solar observations, e.g. pointing parallel 
with an imaging X-ray telescope. One obvious drawback of thicker filters would be the increase of 
the lower operational energy limit of SCXM. These full disc simulation results are also compiled 
in Table 2.\\ 
\hspace*{0.5cm} These kind of full disc solar coronal X-ray observations have been made by SphinX \cite{SPHINX2}. 
SphinX instrument had a very wide dynamical range, because it incorporated three separate directly 
toward the Sun pointing Si PIN detectors, which each had a different effective area. The FOVs of SphinX 
detectors were circular with angular diameters of about 2$^{\circ}$. SphinX was able to made accurate plasma 
diagnostics and precise photometry at an extreme low activity of the Sun. SphinX was also able to measure 
solar X-ray fluxes 100 times weaker than GOES A-level and up to X30 level flux. SCXM would be even more 
sensitive due the its higher effective area with an energy resolution of 250 eV at 6 keV, while observing 
Sun at a quiescent phase. SCXM would be able to perform a diverse solar coronal plasma diagnostics with the 
time resolution of 1 second corresponding to the performance of the existing read out electronics available. 
The practical lower energy limit of SCXM is not dictated by the filters. The noise free low energy limit of 
our existing read out electronics is about 1.5 keV. Hence the Al K$\alpha$ line would be just below the lower 
energy range of this instrument related to the line spectroscopic performance. It could be also possible to 
resolve the locations of the highly ionized iron lines around 6.5 keV or above during the onset of flare 
events above C-level. This same analysis method would be also valid for other ionized coronal elements like 
S, K, Ca and Ar and determination of the respective elemental abundances in full disk observation mode. It 
would also be able to determine the temporal evolution of a plasma temperature and emission measure with 
a time resolution of 1 second corresponding to a photon statistics of 1\% at B-level activity. SCXM would 
be able to do very accurate photometry between 1.5 and 10 keV and reveal the micro-flare variability and 
possible oscillations related to the low activity X-ray emission. SCXM could also derive and confirm the 
new soft X-ray reference photometric standards corresponding to the extreme low activity of the Sun like 
SphinX team has already considered. These levels could represent fluxes of 0.1A (1 $\times 10^{-9}$ Wm$^{-2}$) and 0.01A 
($10^{-10}$ Wm$^{-2}$), which are one and two magnitudes below the sensitivity of GOES detectors.      

\begin{figure}
   \begin{center}
    \epsfig{file=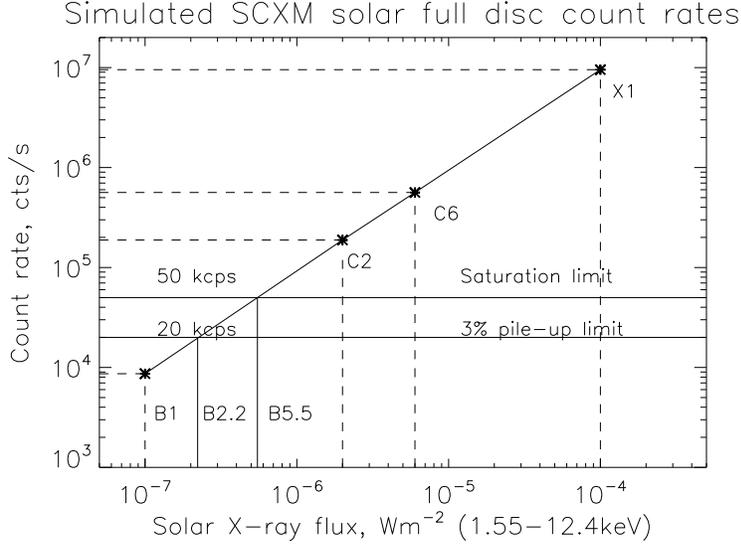,scale=0.60,keepaspectratio}
    \label{Fig:fig5}
    \end{center}
\caption{Simulated SCXM solar full disc count rates for several flux levels.}
\end{figure}

\section{Off-solar sensitivity simulations}
\label{Off-solar sensitivity simulations}

During a high solar activity the hot corona extends significantly. This is clearly seen by 
comparing, e.g. white light photographs taken during total solar eclipses at a high and a quiet 
solar activity. Off-solar coronal observations have been made at EUV band up to the distance of two
solar radii from the center of the Sun \cite{Slemzin}. Similar observations have not been made at 
soft X-ray band, which are obtained solely from the outer solar corona including the full spectral 
information. The X-ray surface brightness can vary significantly around the Sun. Above an active 
region, a coronal mass ejection or a huge flare the coronal X-ray emission can be several orders of 
magnitude greater than on the opposite side of the Sun \cite{Claire}.\\ 
\hspace*{0.5cm} The lack of good spectroscopic off-solar X-ray data, i.e. observations of the X-ray emission coming 
solely off the solar limb from the outer corona was the primary driver to start the designing of 
SCXM instrument concept. SMART-1 XSM already verified that this kind of off-solar observation is 
possible, even XSM was used for full solar limb observations with the large detector FoV. The corresponding 
data analysis of two off-solar spectra are introduced in \cite{Alha2}. These data were obtained 
during X- and M-flare events while the Sun was totally out of the detector FoV. The both analyzed 
spectra included several emission lines above a weak continuum. Since the XSM was not designed for 
such solar observations, there were a possibility also instrument related emission lines in the 
spectrum, probably due to scattering of solar X-rays from outside the FoV. 
 
\subsection{Background estimation}
\label{Background estimation}

The first task was to estimate the background flux levels while simulating weak sources like off-solar 
X-ray surface brightness of an extended solar corona or an axion X-ray decay emission introduced in 
the following section. All these simulations were assumed to be background dominated.

\subsubsection{Sky background model}
\label{Diffuse sky background model}

The diffuse X-ray background spectrum used in these simulations was derived from the data obtained 
by SMART-1 XSM. XSM made several long duration observations in which the Sun was well outside of the 
detector FoV. We have made a rough X-ray background spectral analysis of these off-solar observations. 
The pointings of XSM during these off-limb observations were all close to the Galactic Plane, which 
included several X-ray point sources. Hence our derived power law spectrum model representing the 
X-ray background spectrum is more intense than the diffuse sky background obtained by other instruments 
from the high galactic latitudes, e.g. \cite{Lumb1}, which is based on data obtained by the XMM Newton 
EPIC/MOS instrument. \\
\hspace*{0.5cm} We derived the representative diffuse sky background model for off-solar and axion simulations by 
folding the flux obtained by the XSM through the SCXM grasp. After scaling the model, the normalization 
factor {\it N} is 9.45 $\times 10^{-4}$ photons cm$^{-2}$s$^{-1}$sr$^{-1}$ at 1 keV and the photon index {\it $\Gamma$} 
is 2.17. The respective flux is 1.66 $\times 10^{-10}$ erg cm$^{2}$s$^{-1}$keV$^{-1}$ between 2 and 8 keV.

\subsubsection{Instrument background model} 
\label{Instrument background model}

The instrument background can be estimated analogously for SCXM by applying the cosmic and solar energetic
particle flux obtained from the study of the XMM-newton EPIC pn-CCD X-ray camera \cite{Lumb2}. The CCD is 
made of silicon, i.e. it is a thin planar semiconductor chip as like the PIN diode of SCXM is. XMM-newton 
is in a highly elliptical orbit around the Earth. The post rejection background flux {\it $F_{i}$} for that 
instrument is reported to be 0.0039 photons cm$^{-2}$s$^{-1}$keV$^{-1}$. The total SiPIN detector surface 
area (area of a cylinder) {\it $A_{tot}$} of SCXM would be 2{\it $A_{det}$}+2{\it $\pi$} {\it $r_{det}$} 
{\it h}. The respective numerical value of {\it $A_{tot}$} is 0.47cm$^{-2}$. This {\it $A_{tot}$} takes 
into account the omnidirectional particle bombardment. Thus the number of photons ({\it $F_{i}$}
{\it $A_{tot}$}{\it $\Delta$E}) would be 0.0147 photons/s in the energy range of 2-8 keV. The scaled 
instrumental background flux {\it $F_{I}$} for SCXM corresponds to 8.8 $\times 10^{-11}$ erg cm$^{-2}$s$^{-1}$keV$^{-1}$ 
between 2 and 8 keV. The instrument background spectrum {\it $F_{I}$} was assumed to be constant at 
each energy. The above instrument background may be over estimated, because SCXM would be equipped with a 
Be filter with a thickness of 13$\mu$m. The thickest filter configuration of the EPIC pn-CCD is only made 
of 0.1$\mu$m layer of Sn and 0.15 $\mu$m layer of Al. Hence SCXM would have a better shielding at least 
against low energy electrons coming through the aperture. On the other hand, all the high energy particles
hitting the detector PIN diode are recorded in the highest nominal energy channel of 20 keV. Hence the 
particle events could be easily distinguished from the solar coronal X-ray spectrum between 1 and 10 keV 
excluding the possible low energy secondary emission. The situation is opposite in off limb and axion 
observations, when the coronal X-ray brightness is expected to be low and observations become background 
dominated. The both background spectra used in the simulations are plotted in Fig. 6.

\begin{figure}
   \begin{center}
    \epsfig{file=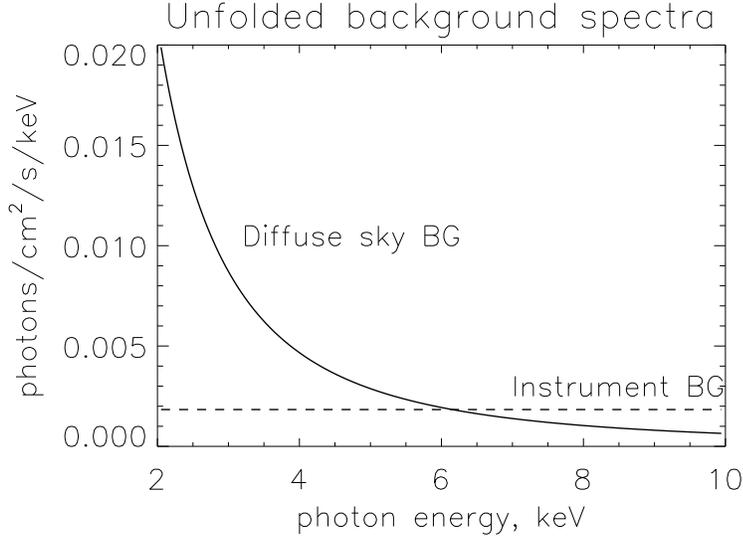,scale=0.60,keepaspectratio}
    \label{Fig:fig6}
    \end{center}
\caption{The solid line in the plot represents the diffuse X-ray background spectrum derived 
              from the SMART-1 XSM off-solar observations. The dashed line represents the applied 
              instrumental background model for SCXM.}
\end{figure}

\subsection{Off-solar coronal plasma diagnostics simulation}
\label{Off-solar coronal plasma diagnostig simulation}

We have simulated the off-solar X-ray emission between 1.55 keV and 12.4 keV with SCXM. We applied a 
bremsstrahlung spectral model in our simulations. This isothermal model composes of two parameters. 
The first parameter is a normalization factor which is related to the emission measure ({\it EM}), 
i.e. the density of the X-ray emitting plasma. The other parameter is the photon index, which relates 
to the plasma temperature. We made two simulations to derive the model parameter errors corresponding 
to 10\%. The integration time of the first simulation was 100 seconds and the other simulation was 
made with an integration time of 1000 seconds. Each simulation included the above background models. 
The plots in Fig. 7 represent the simulation results. The 100 s curve corresponds roughly 
to a flux of $10^{-10}$ Wm$^{-2}$/FoV and the 1000 s integration corresponds to $10^{-11}$ Wm$^{-2}$. The 
lowest quiescence GOES A-level flux is roughly $10^{-8}$ Wm$^{-2}$. Hence SCXM could be able to detect 
the coronal X-ray emission between 1.55 keV and 12.4 keV within 100 seconds within $\pm10\%$ error 
limits of the model parameters, when the flux is one hundredth of the A-level.\\  
\hspace*{0.5cm} This simple simulation shows that SCXM could be an suitable instrument for spectroscopically map the 
spatial distribution of off-solar coronal X-rays. During a high solar activity, e.g. a flare occurring 
on the edge of the solar limb may cause enhanced X-ray emission far out of the Sun. These kind of 
observations would be the primary scope of SCXM in an off-solar mode, i.e. observing X-ray emission 
from the expanding corona during a large CMEs and flares. These observations would be able to reveal, 
how far the hot X-ray emitting plasma can extend in the solar corona. SCXM would roughly map 
the coronal plasma parameters, i.e. emission measure and temperature as a function of elongation angle 
measured off the Sun.

\begin{figure}
   \begin{center}
    \epsfig{file=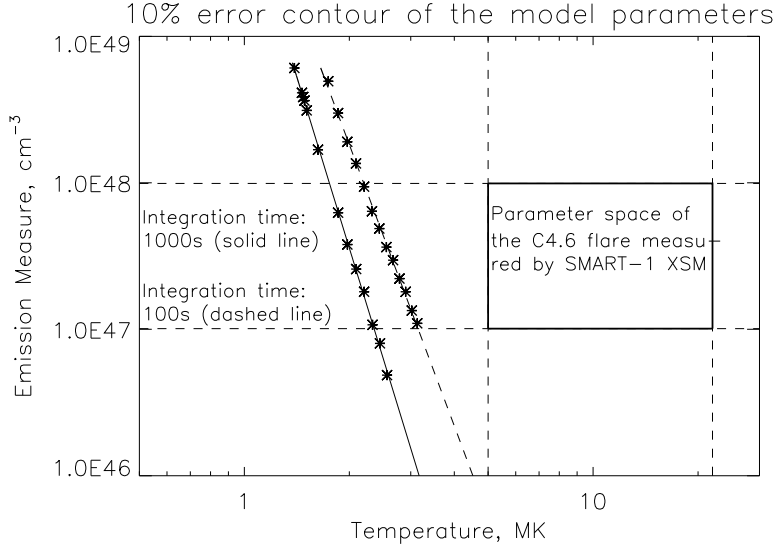,scale=0.60,keepaspectratio}
    \label{Fig:fig7}
    \end{center}
\caption{The solid line in the plot represents simulations corresponding to the integration time of 1 ks 
              and the dashed line represents simulations corresponding to the integration time of 100 s. The
              asterisks signs represent the simulated data points, where the errors of the EM and temperature
              were less than 10\%. The rectangular box on the right represents the parameter space of of the EM 
              and T measured by SMART-1 XSM during a C-level flare.}
\end{figure}

\subsection{Hypothetical axions X-ray emission simulations during a solar quiescence}
\label{Hypothetical axions X-ray emission simulations during a solar quiescence}

The purpose of this section is to introduce the new instrument concept under consideration 
to experimental particle physicists to use SCXM in a space borne axion X-ray observations. 
Several research groups are working with a ground-based and underground instrumentation aiming to 
find an X-ray signature of decaying axions, e.g. the {\it CAST} experiment at CERN \cite{Aalseth} 
and the {\it CDMS} experiment at the Soudan Underground Laboratory \cite{Ahmed}. A more detailed 
treatment about the fundamentals of solar related axion physics can be found, e.g. in \cite{Carlson}, 
\cite{Zioutas} and \cite{DiLella1}.

\subsubsection{Off-solar axion spectrum simulation}
\label{Off-solar axion spectrum simulation}

The applied axion spectra used in these simulations were obtained from the paper \cite{DiLella2}. 

\begin{equation}
F_{S}=K(E/1keV)^{-\Gamma} e^{-E/E_{co}}
\end{equation}

This spectrum represents massive Kaluza-Klein axions, which are gravitationally trapped in
the proximity of the Sun. These axions are assumed to orbit around the Sun by forming a halo
like object, where the particle density of axions decreases radially outward from the Sun 
obeying an exponential law. (See Fig. 8). These simulated flux values correspond 
to the respective pointings of SCXM, i.e. the surface brightness at a distinct elongation angle.

\begin{figure}
   \begin{center}
    \epsfig{file=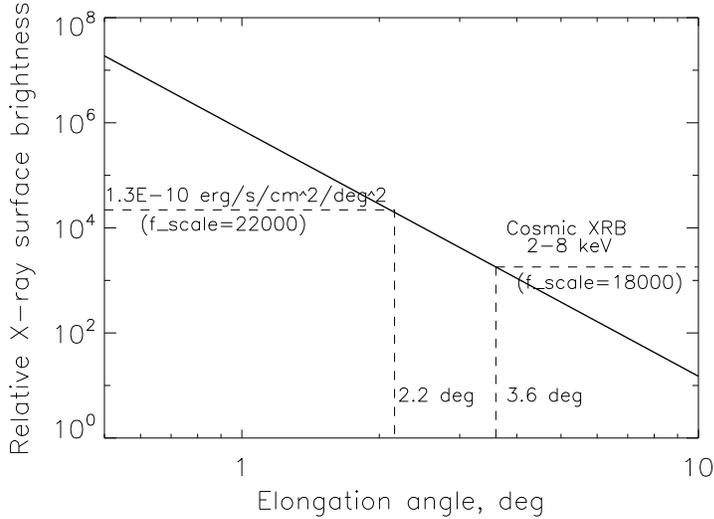,scale=0.6,keepaspectratio, angle=0}
    \label{Fig:fig8}
    \end{center}
\caption{A relative X-ray surface brightness of gravitationally trapped axions 
              as a function of elongation angle from the Sun. The cosmic X-ray background flux at 
              about 3.6$^{\circ}$ is 18000 (relative scale) corresponding to 1.1 $\times 10^{-11}$ erg s$^{-1}$cm$^{-2}$
              deg$^{-2}$keV$^{-1}$ (2 keV -8 keV) according to \cite{Moretti}. This connection enables 
              to estimate all the surface brightness fluxes in physical units with respect to the 
              elongation angle. The over drawn horizontal black solid line represents the total background
              flux level used in our simulations. The respective limiting $\theta$ is about 2.2$^{\circ}$.
              This plot is redrawn with the authors' permission from \cite{Zioutas}.}
\end{figure}

We have omitted the possible extended solar dynamical and scattering effects related to the coronal 
X-ray emission in these simulations, i.e. we have assumed a quiescent solar phase. The observational 
constraints were only depending on the background fluxes. (See Section 4.1)\\ 
\hspace*{0.5cm} The theoretical axion flux was obtained from \cite{DiLella2} and the spectral model was approximated 
with a cutoff-power law spectral model shown in Eq. 1. This model includes three free parameters. The 
first model parameter denotes the photon index {\it $\Gamma$}. The second parameter denotes the cutoff 
energy {\it E$_{co}$ } and  the last parameter {\it K} denotes the normalization. This model approximates 
very accurately the theoretically calculated surface brightness of decaying axions around the Sun. The 
authors in the paper \cite{DiLella1} have used a similar approximation representing the spectrum of 
decaying axions. The following parameter set was used: {\it $\Gamma$} was -3, {\it $E_{co}$} was 1.54 
keV and {\it K} was a free parameter used for scaling the simulated flux at different elongation angles 
$\theta$. An unfolded and folded axion spectra obtained at $\theta$ of 2$^{\circ}$ are shown in 
Fig. 9.\\
\hspace*{0.5cm} We made spectral simulations of the X-ray emission of decaying axions at elongation angles of 
3$^{\circ}$, 3.5$^{\circ}$, 4$^{\circ}$, 5$^{\circ}$, 5.5$^{\circ}$ and 6$^{\circ}$ to determine the sensitivity 
of SCXM as a function {\it $\theta$}. The applied surface brightness and the corresponding normalization 
parameter were determined from the plot shown in Fig.8. Then the spectral models were folded through the 
SCXM's grasp and RMF with different integration times to derive the respective signal-to-noise ratios 
{\it SNR}. {\it SNR} describes the significance of the detections and they were calculated with the 
formula related to a background dominated observation shown in Eq. 2 \cite{Fraser}. The parameters of 
the formula are listed in Table 3. The observation geometries of these axion simulations are illustrated 
in Fig. 10.

\begin{figure}
   \begin{center}
    \epsfig{file=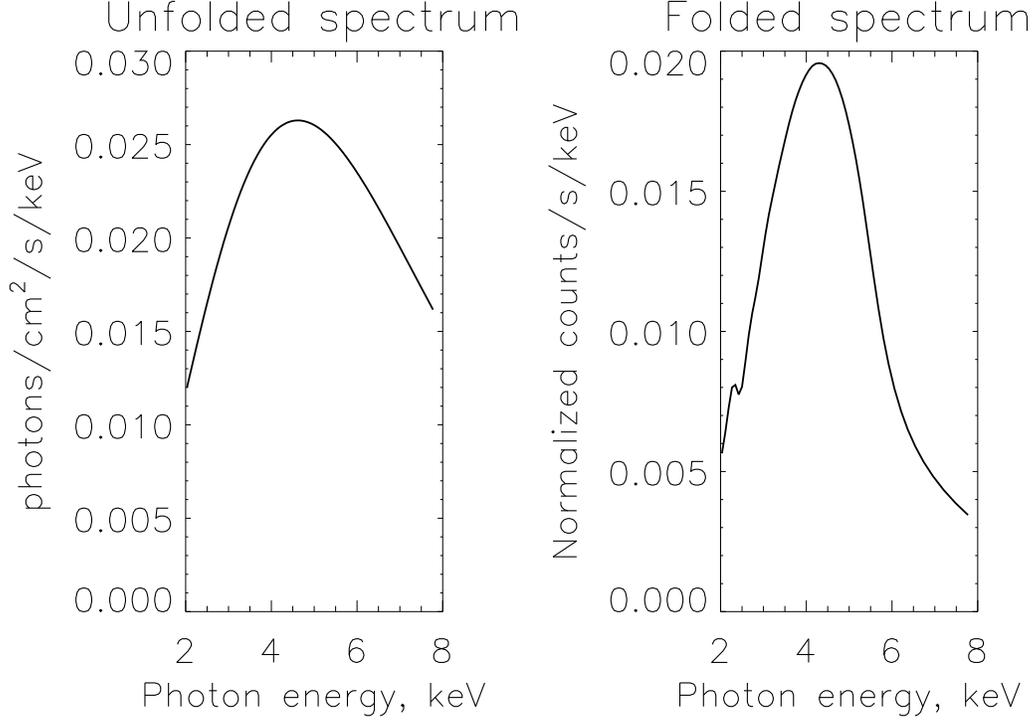,scale=0.8,keepaspectratio, angle=0}
    \label{Fig:fig9}
    \end{center}
\caption{The plot on the left side represents the unfolded axion spectrum observed 
              2$^{\circ}$ off the solar disc center. The plot on the right side represents the same 
              axion spectrum folded through SCXM. Integration time was 15 ks.}
\end{figure}

The plots in Fig. 11 represent the SCXM axion simulation results obtained at different 
$\theta$, where each of the simulated SNR is plotted as a function of the integration time. The plot 
in Fig. 12 illustrates the relation between minimum detectable flux at 3$\sigma$ detection 
significance as a function of $t_{int}$. 

\begin{equation}
SNR=\frac{GF_{S}\sqrt t_{int}}{\sqrt{\Delta E(F_{I}A_{tot}+GF_{SKY})}}
\end{equation}

\begin{figure}
   \begin{center}
    \epsfig{file=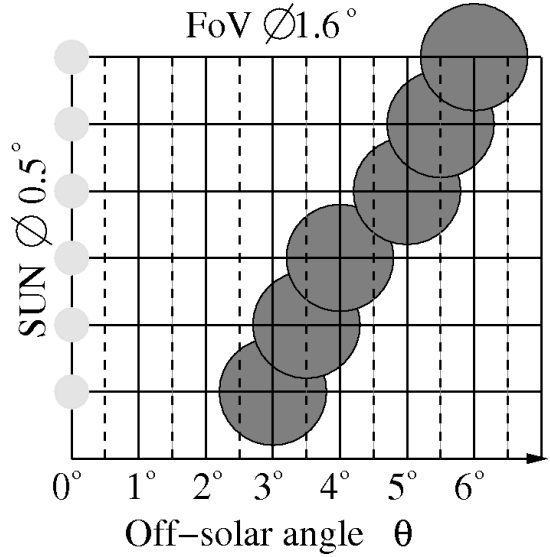,scale=0.4,keepaspectratio, angle=0}
    \label{Fig:fig10}
    \end{center}
\caption{Angular scales of the observation geometries of the axion simulations.}
\end{figure}

\begin{figure}
   \begin{center}
    \epsfig{file=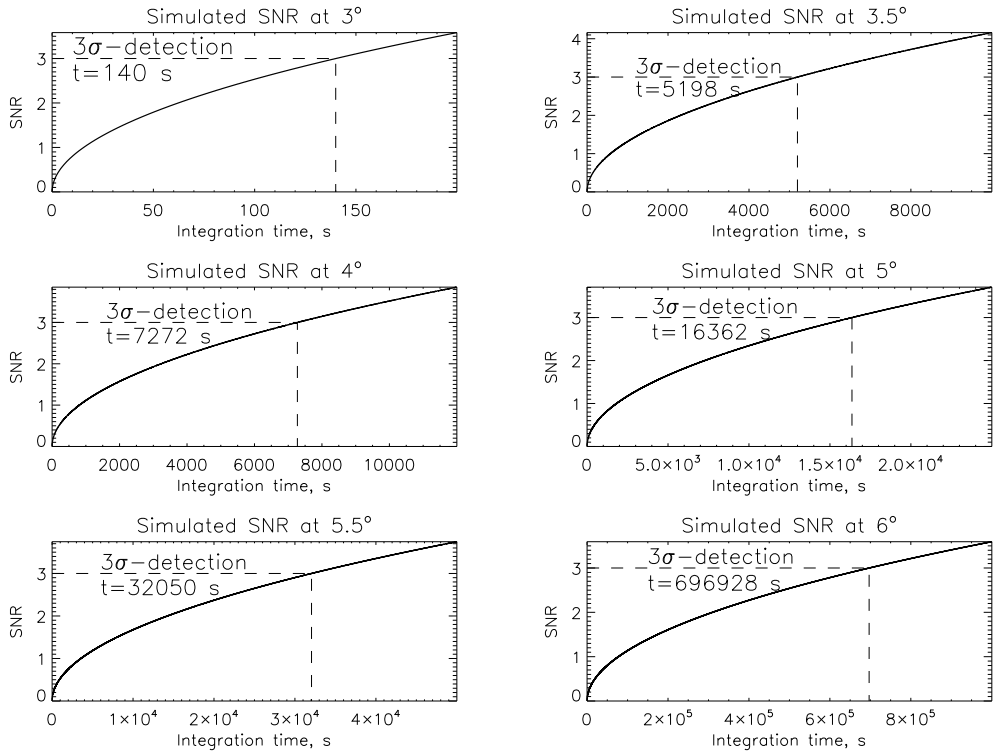,scale=0.8,keepaspectratio, angle=0}
    \label{Fig:fig11}
    \end{center}
\caption{The above curves illustrate the SNR as a function of the integration time derived at elongation 
              angles of 3$^{\circ}$, 3.5$^{\circ}$, 4$^{\circ}$, 5$^{\circ}$, 5.5$^{\circ}$ and 6$^{\circ}$. The dashed lines 
              correspond to the integration time required to achieve a 3$\sigma$ significance in the simulated observation. 
              The initial values and the respective simulation results are also compiled in Table 4.}
\end{figure}

\begin{figure}
   \begin{center}
    \epsfig{file=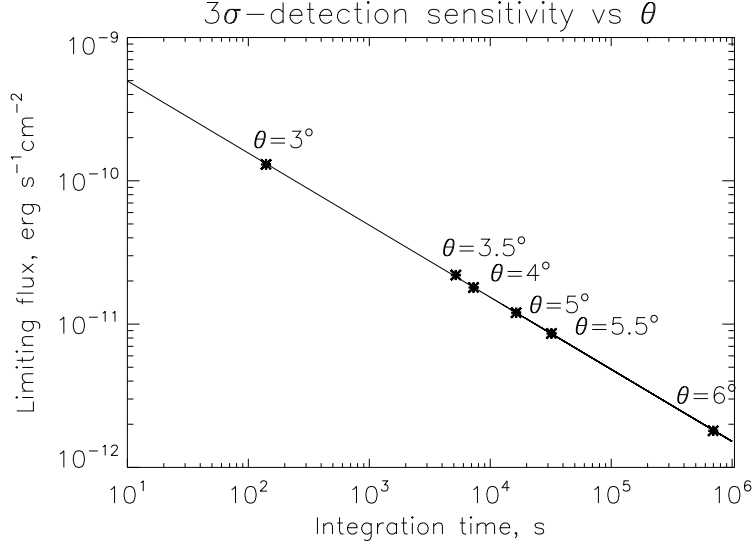,scale=0.6,keepaspectratio, angle=0}
    \label{Fig:fig12}
    \end{center}
\caption{The simulated 3$\sigma$ detection sensitivity of SCXM. The asterisk signs denote the data 
              points corresponding to the different elongation angles $\theta$. The horizontal axis represents 
              the required integration time.}
\end{figure}

These rough broad band continuum simulations encourage to test the usefulness of this hypothetical SCXM 
instrument in space borne observations to find the X-ray signature of decaying axions surrounding the 
Sun. These simulation results shows that SCXM would be sensitive enough to observe X-ray emission from 
these hypothetical particles surrounding the Sun whit in a reasonable integration time up to 5$^{\circ}$.\\

\section{SCXM telemetry budget}
\label{SCXM telrmetry budget}

The telemetry budget of SCXM is estimated according to the design of the SIXS X-ray spectrometer 
\cite{Huovelin}, which will be a payload instrument on board the BepiColombo satellite mission to 
Mercury. The telemetry rate depends on the integration and housekeeping HK sampling time. The estimated 
telemetry rate is compiled in Table 5. Two different versions of the readout 
electronics is considered. The first option includes 512 energy channels and covers a nominal energy 
range of 1-20 keV. The other option includes only 256 energy channels and covers a nominal energy 
range of 1-10 keV. The nominal channel width in the both scales is around 40 eV/ch. There would be 
no need to cover the energy range above 10 keV, because the solar coronal continuum and line emission 
and the respective plasma diagnostics of interest lies mostly between photon energies of 1 keV and 
10 keV. This same criterion is also valid for axion-photon emission for which the maximum flux is 
estimated to be around 4.5 keV. Each of the spectrum channel corresponds to 4 bytes. The data size of 
the HK depends on the number of the monitored parameters, e.g. detector PIN and ambient box temperatures,
leakage current, several voltage and currents. A good approximation of the byte size for this kind of 
instruments is about 64 bytes including header sizes and all required extra parameters, e.g. time 
stamps. Hence this will contribute in maximum 32 bytes. The final telemetry rate is linearly proportional
to the the collection time of spectra and HK. The plot in Fig. 13 illustrates the data 
rate versus integration time ({\it $t_{int}$}).     

\begin{figure}
   \begin{center}
    \epsfig{file=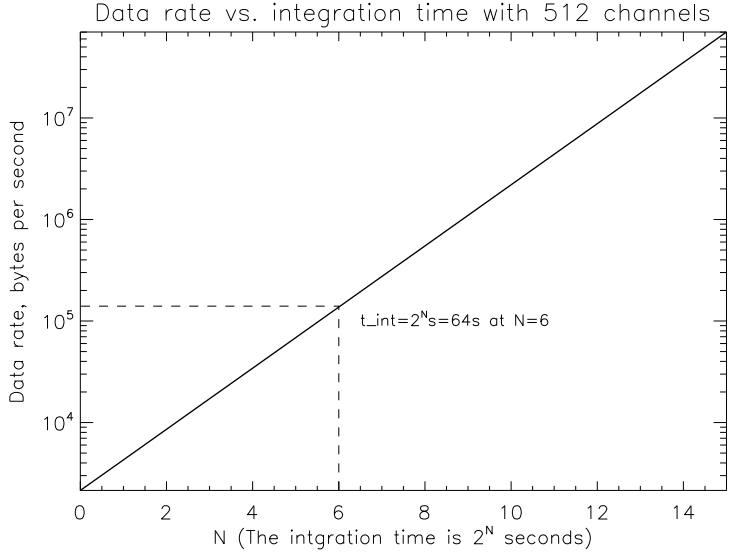,scale=0.60,keepaspectratio}
    \label{Fig:fig13}
    \end{center}
\caption{Total data rate of SCXM equipped with 512 energy channels as a function of $t_{int}$.
              SCXM will be a low level load with respect to the data rate even without of compression.}
\end{figure}

\section{Conclusions}
\label{Conclusions}

Two different observation scenarios are studied by folding models through the SCXM instrument, 
which is under a design phase. This instrument is a very rapid X-ray spectrometer operating 
together e.g. with an broad band Wolter-1 type telescope. SCXM would enhance significantly the mutual 
spectral data analysis with respect to full disc observations when working parallel and with an other 
imaging instrument system of lower energy resolution. In this scenario both instruments would be 
co-aligned on-board a spinning 1D- or 3D-stabilized satellite platform. This would mean that the 
SCXM quantum efficiency should be reduced by applying thicker filters to decrease the count rate level. 
The drawback would be an increment of the lower energy limit of SCXM. The {\it QE} of the instrument 
at lower energies is very sensitive to the choice of filter materials and respective thicknesses. It 
would also be possible to equip SCXM with a rotational multifilter wheel to enable a wider dynamical 
range with respect to the solar activity. One section of this filter wheel would contain also the 
required in-flight $^{55}$Fe calibration source. The previous XSMs were equipped with a bistable 
shutter mechanism, which included the in-flight calibration source. If SCXM is provided with a filter 
wheel, it would cause major mechanical design changes related to the existing XSM technology. It would 
be better to reduce the grasp by changing the optical design of SCXM in the case on-solar sole 
observation mode. On the other hand, a special task of SCXM could be observations of solar X-ray 
emission at solar quiescence below GOES A-level. In this case the current effective area would be 
optimal for that purpose to resolve weak emission, e.g. with a time resolution of one second. The 
solar coronal soft X-ray emission has been observed only by the Sphinx PIN spectrometer in 2009 during 
a grand minimum. To perform similar X-ray photometry during a low activity requires a very sensitive 
and fast detector concept \cite{SPHINX3}. The schematics of the SCXM system is illustrated in 
Fig. 14.\\
\hspace*{0.5cm} The second possible operational scenario would be the off-limb pointing observations, e.g. to observe 
X-ray emission from expanded solar coronal plasma during high activity phase or hypothetical axion
decay emission during solar quiescence. This kind of observation would be the 
first space borne experiment to search for a signature from decaying axions in the halo around the Sun.
The introduced axion X-ray spectral simulations verified that SCXM could resolve the axion signal with 
3$\sigma$ level above the background in reasonable integration time up to elongation angles $\theta$
of 6$^{\circ}$ excluding the solar coronal X-ray emission itself.

\begin{figure}
   \begin{center}
    \epsfig{file=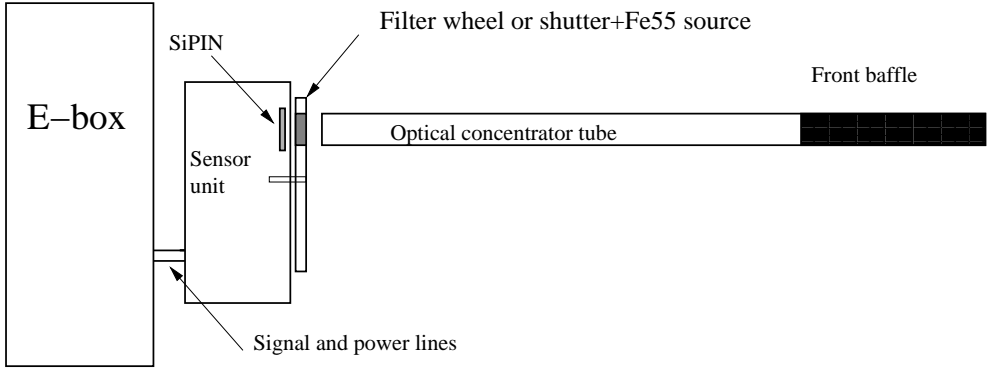,scale=0.70,keepaspectratio, angle=270}
    \label{Fig:fig14}
    \end{center}
\caption{Schematics of the main components of the SCXM instrument. (The dimensions are not in scale.}
\end{figure}

\section{Acknowledgments}
\label{Acknowledgements}

We would like to express our gratitude for the staff of Oxford Instruments Analytical Inc. in Espoo, 
Finland, which has provided us Si PIN technology for space missions and advice for understanding its 
usage and performance. Special thanks to several colleagues at the University of Helsinki for their 
editorial help.\\
We are also very grateful for the advice given by prof. Konstantin Zioutas, who has helped us with
the axion related issues.

\clearpage

\renewcommand{\arraystretch}{0.75}
\begin{table}
\begin{center}
\begin{tabular}{|l|l|}
\hline
Material Component     & Nominal value \\ 
\hline
Si detector thickness {\it h} & 500.0 $\mu$m\\ 
\hline
Dead layer, Si         & 0.01 $\mu$m \\ 
\hline
Cathode and thermal filter, Al & 0.60 $\mu$m \\ 
\hline
Polyimide thermal filter  & 0.25 $\mu$m  \\ 
\hline
Be-filter              & 13.0 $\mu$m  \\ 
\hline
Active detector diameter {\it $d_{det}$} & 5.0 mm   \\ 
\hline
Outer mirror tube aperture diameter $D_{2}$ & 15.0 mm   \\ 
\hline
Outer mirror tube aperture diameter $D_{1}$ & 12.4 mm   \\ 
\hline
Inner mirror tube aperture diameter $d_{2}$ & 10.4 mm   \\ 
\hline
Inner mirror tube aperture diameter $d_{1}$ & 6.0 mm   \\ 
\hline
Concentrator total length $L$ & 179.0 mm   \\ 
\hline
Concentrator outer mirror length $L_{1}$ & 95.0 mm   \\ 
\hline
Concentrator inner mirror length $L_{2}$ & 84.0 mm   \\ 
\hline
Mirror material & Au            \\ 
\hline
Grazing angle & 0.8$^{\circ}$  \\ 
\hline

\end{tabular}
\end{center}
\caption{{\it The nominal dimensions of the detector, filters and optics of SCXM.}}
\label{Table:Table1}
\end{table}

\renewcommand{\arraystretch}{0.75}
\begin{table}
\begin{center}
\begin{tabular}{|l|l|l|l|l|}
\hline
GOES scale   & Flux, Wm$^{-2}$ & T, keV (MK) & Norm, (EM, cm$^{-3}$)  & cts/s\\
\hline
B1 & 1.0 $\times 10^{-7}$ & 0.42189 (4.9)  & 4.15100 $\times 10^{-6}$ (3.9 $\times 10^{48}$) & 3.43100 $\times 10^{3}$\\
\hline
C2 & 2.0 $\times 10^{-6}$ & 0.98443 (11.4) & 6.11805 $\times 10^{-6}$ (5.7 $\times 10^{48}$) & 1.88314 $\times 10^{5}$\\
\hline
C6 & 6.0 $\times 10^{-6}$ & 1.54268 (17.9) & 6.90000 $\times 10^{6}$ (6.5 $\times 10^{48}$) & 5.63164 $\times 10^{5}$\\
\hline
X1 & 1.0 $\times 10^{-4}$ & 1.69276 (19.6) & 1.10000 $\times 10^{8}$ (1.0 $\times 10^{50}$) & 9.53901 $\times 10^{6}$\\
\hline
\end{tabular}
\end{center}
\caption{{\it The on-axis (full limb) solar simulation results.}}
\label{Table:Table2}
\end{table}

\renewcommand{\arraystretch}{0.75}
\begin{table}
\begin{center}
\begin{tabular}{|l|l|l|}
\hline
Parameter & Meaning & Value and Dimension\\
\hline
SNR & Signal-to-noise & dimensionless (See Fig. 11)\\
\hline
G & Grasp & cm$^{2}$deg$^{2}$ (See Fig. 4)\\
\hline
$F_{S}$ & 3$\sigma$ source flux &  erg s$^{-1}$cm$^{-2}$deg$^{-2}$keV$^{-1}$ (See column 3 of Tab. 3)\\
\hline
$F_{I}$ & Instrument flux & $8.8 \times 10^{-11}$ erg s$^{-1}$cm$^{-2}$keV$^{-1}$ (2-8 keV)\\
\hline
$F_{SKY}$ & Diffuse background flux & $7.9 \times 10^{-11}$ erg s$^{-1}$cm$^{-2}$deg$^{-2}$keV$^{-1}$ (2-8 keV)\\
\hline
$A_{tot}$ & Total detector area & 0.47 cm$^{2}$\\
\hline
$t_{int}$ & Integration time & s (Variable)\\
\hline
$\Delta$E & Bandwidth & 2 keV-8 keV \\
\hline
\end{tabular}
\end{center}
\caption{{\it Description of the parameters in the formula for deriving a background dominated SNR.}}
\label{Table:Table3}
\end{table}

\renewcommand{\arraystretch}{0.75}
\begin{table}
\begin{center}
\begin{tabular}{|l|l|l|l|l|l|l|}
\hline
$\theta$, $^{\circ}$ & $f_{scale}$ & Flux, erg s$^{-1}$cm$^{-2}$deg$^{-2}$keV$^{-1}$ (2-8 keV) & count rate, 1/s & $t_{int}$ @ 3$\sigma$, s & Normalization \\
\hline
3   & 23.45 & $6.2 \times 10^{-11}$ & $8.9 \times 10^{-3}$ & 140 & $6.7 \times 10^{-4}$\\
\hline
3.5 & 2.01  & $1.0 \times 10^{-11}$ & $1.5 \times 10^{-3}$ & 5198 & $1.1 \times 10^{-4}$\\
\hline
4   & 1.68  & $8.6 \times 10^{-12}$ & $1.2 \times 10^{-3}$& 7272 & $9.3 \times 10^{-5}$\\
\hline
5   & 1.12  & $5.7 \times 10^{-12}$ & $8.3 \times 10^{-4}$ & 16362 & $6.2 \times 10^{-5}$\\
\hline
5.5 & 0.78  & $4.1 \times 10^{-12}$ & $5.9 \times 10^{-4}$ & 32050 & $4.4 \times 10^{-4}$\\
\hline
6   & 0.17  & $8.6 \times 10^{-13}$ & $1.3 \times 10^{-4}$& 696928 & $9.5 \times 10^{-6}$\\
\hline
\end{tabular}
\end{center}
\caption{{\it Off-solar axion simulation results. The parameter $f_{scale}$ was needed to convert the
              surface brightness into physical units in the second column. The third column includes 
              the respective fluxes used in the simulation. The model predicted count rates are in 
              the fourth column. 3$\sigma$ integration times are in the fifth column. The normalization
              parameters of the used cut-off power law spectrum model are in the last column. All
              the simulations were made at the energy range $\Delta$E between 2 keV and 8 keV.}}
\label{Table:Table4}
\end{table}

\renewcommand{\arraystretch}{0.55}
\begin{table}
\begin{center}
\begin{tabular}{|l|l|}
\hline
\textbf{Data source}              & \textbf{Nominal data size} \\ 
\hline
Spectrum of 512 channels & 2048              \\
\hline
Spectrum of 256 channels & 1024              \\
\hline
{\it HK}                 & 64                \\
\hline
Header                   & 16                \\
\hline
Others                   & 16                \\
\hline
\end{tabular}
\end{center}
\caption{{\it The contribution of the different components of SCXM to the estimated telemetry rate.}}
\label{Table:Table5}
\end{table}


\clearpage

\begin{thebibliography}{00}

\bibitem{Aalseth}    C. E. Aalseth, E. Arikc, D. Autiero, F. T. Avignone, K. Barth, S. M. Bowyer, H. Brauninger,
                     L. Brodzinski, L. Brodzinski, S. Cebrian, G. Celebi, S. Cetin, J. I. Collar, R. Creswick,
                     A. Delbart, M. Delattre, L. DiLella, R. De Oliveira, Ch. Eleftheriadis, N. Erdutan, G. Fanourakis,
                     H. A. Farach, C. Fiorini, Th. Geralis, I. Giomataris, T. A. Girard, S. N. Gninenko, N. A. Golubev,
                     M. Hasinoff, D. Hoffmann, I. G. Irastorza, f, J. Jacoby, F. Jeanneau, M. A. Knopf, A. V. Kovzelev, 
                     R. Kotthaus, M. Krcmar, Z. Krecak, B. Lakic, A. Liolios, A. Ljubicic, G. Lutz, A. Longoni, G. Luzon, 
                     A. Mailov, V. A. Matveev, H. S. Miley, A. Morales, J. Morales, M. Mutterer, A. Nikolaidis, S. Nussinov, 
                     A. Ortiz, W. K. Pitts, A. Placci, V. E. Postoev, G. G. Raffelt, H. Riege, M. Sampieto, M. Sarsa, 
                     I. Savvidis, M. Stipcevic, C. W. Thomas, R. C. Thompson, P. Valco, J. A. Villar, B. Villierme, 
                     L. Walckiers, W. Wilcox, K. Zachariadou and K. Zioutas, The Cern Axion Solar Telescope (CAST),
                     Nuclear Physics B (Proc. Suppl.) 110 (2002) p. 85-87 
\bibitem{Ahmed}      Z. Ahmed, D. S. Akerib, S. Arrenberg, C. N. Bailey, D. Balakishiyeva, L. Baudis, D. A. Bauer, J. Beaty,
                     P. L. Brink, T. Bruch, R. Bunker, B. Cabrera, D. O. Caldwell, J. Cooley, P. Cushman, F. DeJongh,
                     M. R. Dragowsky, L. Duong, E. Figueroa-Feliciano, J. Filippini, M. Fritts, S. R. Golwala, D. R. Grant, J. 
                     Hall, R. Hennings-Yeomans, S. Hertel, D. Holmgren, L. Hsu, M. E. Huber, O. Kamaev, M. Kiveni, M. Kos,
                     S.W. Leman, R. Mahapatra, V. Mandic, D. Moore, K. A. McCarthy, N. Mirabolfathi, H. Nelson,
                     R.W. Ogburn, M. Pyle, X. Qiu, E. Ramberg, W. Rau, A. Reisetter, T. Saab, B. Sadoulet, J. Sander,
                     R.W. Schnee, D. N. Seitz, B. Serfass, K. M. Sundqvist, M. Tarka, G. Wang, S. Yellin, J. Yoo and 
                     B. A. Young, Search for Axions with the CDMS Experiment, PRL 103, 141802 (2009)
\bibitem{Alha1}      L. Alha a, J. Huovelin, T. Hackman, H. Andersson, C.J. Howe, Eero Esko, M. Va\"a\"an\"anen, The 
                     in-flight performance of the X-ray Solar Monitor(XSM) on-board SMART-1, Nuclear Instruments and Methods 
                     in Physics Research A 596 (2008) p. 317-326
\bibitem{Alha2}      L. Alha, Off-solar X-ray observations and a new detector concept with a concentrator optics, Nuclear 
                     Instruments and Methods in Physics Research A (2009) p. 497-504 
\bibitem{Arnaud}     K. A. Arnaud,, 1996, Astronomical Data Analysis Software and
                     Systems V, eds. Jacoby G. and Barnes J., p17, ASP Conf. Series
                     volume 101.
\bibitem{Carlson}    Eric D. Carlson and Li-Sheng Tseng, Pseudoscalar conversion and X-rays from the Sun, Physics Letters B
                     Volume 365, Issues 1-4, (1996) 193-201
\bibitem{Claire}     Claire L. Raftery1, Peter T. Gallagher, R. T. James McAteer, Chia-Hsien Lin, and Gareth Delahunt, EVIDENCE 
                     FOR INTERNAL TETHER-CUTTING IN A FLARE/CME OBSERVED BY MESSENGER, RHESSI AND STEREO, The Astrophysical 
                     Journal, (2010) 721:1579-1584
\bibitem{DiLella1}   L. Di Lella, A. Pilaftsis, G. Raffelt and K. Zioutas, Search for solar Kaluza-Klein axions in theories 
                     of low-scale quantum gravity, PHYSICAL REVIEW D, VOLUME 62, 125011 (2000)  
\bibitem{DiLella2}   L. DiLella and K. Zioutas, Observational evidence for gravitationally trapped massive axion(-like) 
                     particles, Astroparticle Physics 19 (2003) 145-170 
\bibitem{Fraser}     G. W. Fraser, X-ray detectors in astronomy, Cambridge University Press 1989 

\bibitem{GOES}       $http://www.ngdc.noaa.gov/stp/GOES/goes_legend.htm$
\bibitem{Huovelin}   J. Huovelin, R. Vainio, H. Andersson, E. Valtonen, L. Alha, A. M\"alkki, M. Grande, G.W. Fraser, M. Kato, 
                     H. Koskinen, K. Muinonen, J. N\"ar\"anen, W. Schmidt, M. Syrj\"suo, M. Anttila, T. Vihavainen, E. Kiuru, 
                     M. Roos, J. Peltonen, J.Lehti, M. Talvioja, P. Portin and M. Pryddercl, Solar Intensity X-ray and 
                     particle Spectrometer (SIXS), Planetary and Space Science Volume 58, Issues 1-2 (2010) 96-107  
\bibitem{Lumb1}      D. H. Lumb, R. S. Warwick, M. Page and A. De Luca, X-ray background measurements with XMM-Newton EPIC,
                     A\&A Volume 389, Number 1 (2002) 93-105 
\bibitem{Lumb2}      D. H. Lumb et al., Proceedings of New Visions of the X-ray Universe in the XMM-Newton and 
                     Chandra Era, ESA SP-488 (2002)
\bibitem{Moretti}    A. Moretti, S. Campana, D. Lazzati, and G. Tagliaferri, The Resolved Fraction of the Cosmic X-Ray Background, 
                     The Astrophysical Journal, 588 (2003) p. 696-703 
\bibitem{Shibata}    K. Shibata and Takaaki Yokoyama, A Hertzsprung-Russel-like diagram for solar/stellar and corona emission measure versus 
                     temperature diagram, Astrophysical Journal, 577, p. 422, 2002
\bibitem{Slemzin}    V. Slemzin, O. Bougaenko, A. Ignatiev, S. Kuzin, A. Mitrofanov, A. Pertsov, and I. Zhitnik, Off-limb EUV 
                     observations of the solar corona and transients with the CORONAS-F/SPIRIT telescope-coronagraph, Ann. Geophys., 
                     26, 3007-3016, 2008        
\bibitem{SPHINX1}    J. Sylwester, , S. Kuzin, Yu. D. Kotov, F. Farnik and F. Reale, SphinX: A Fast Solar Photometer in X-rays, 
                     J. Astrophys. Astr. (2008) 29, 339-343
\bibitem{SPHINX2}    S. Gburek, J. Sylwester, M. Kowalinski, J. Bakala, Z. Kordylewski, P. Podgorski, S. Plocieniak, M. Siarkowski, 
                     B. Sylwester and W. Trzebinski, et al., SphinX soft X-ray spectrophotometer: Science objectives, design and 
                     performance, SSN 0038 0946, Solar System Research, 2011, Vol. 45, No. 3, pp. 189-199. © Pleiades Publishing, 
                     Inc., 2011.
\bibitem{SPHINX3}    J. Sylwester, M. Kowalinski, S. Gburek, M. Siarkowski, S. Kuzin, F. Farnik, F. Reale  Phillips and J.H. Kennth,
                     The Sun's X-ray Emission During the Recent Solar Minimum. Eos, Transactions American Geophysical Union, 
                     Volume 91, Issue 8, p. 73-74 (2010)
\bibitem{Zioutas}    K. Zioutas, K. Dennerl, L. DiLella, D. H. H. Hoffmann, J. Jacoby, and Th. Papaevangelou, Quiet-Sun X-Rays as 
                     Signature for New Particles, Astrophysical Journal 607 (2004), p. 575-579
\bibitem{YOKHOH}     P. A. Sturrock, M. S. Wheatland and L. W. Acton, Yokhoh Soft X-Ray Telescope Images of the Diffuse Solar Corona, 
                     Astrophysical Journal 461 (1996), p. 115-117 

\end{thebibliography}
\end{document}